\documentclass[12pt]{article}

\usepackage{graphics}
\usepackage{graphicx, subfigure}
\usepackage{amssymb}
\usepackage{amsmath}

\newcommand{\rf}[1]{(\ref{#1})}
\newcommand{\beq}{\begin{equation}}
\newcommand{\eeq}{\end{equation}}
\newcommand{\bea}{\begin{eqnarray}}
\newcommand{\eea}{\end{eqnarray}}

\newcommand{\e}{\mbox{e}}

\newcommand{\g}{\gamma}

\newcommand{\Lam}{\Lambda}
\newcommand{\bt}{\beta}

\newcommand{\n}{\nu}
\newcommand{\m}{\mu}

\newcommand{\tht}{\theta}
\newcommand{\ep}{\varepsilon}

\newcommand{\del}{\delta}
\newcommand{\Del}{\Delta}

\newcommand{\oh}{\frac{1}{2}}
\newcommand{\oq}{\frac{1}{4}}

\newcommand{\tr}{\mathrm{tr}\,}
\newcommand{\ra}{\rangle}
\newcommand{\la}{\langle}
\newcommand{\prt}{\partial}

\newcommand{\cT}{{\cal T}}

\newcommand{\hT}{{\hat{T}}}
\newcommand{\hH}{{\hat{H}}}

\newcommand{\hE}{{\hat{E}}}
\newcommand{\hP}{{\hat{P}}}

\usepackage[american]{babel}
\usepackage[latin1]{inputenc}
\usepackage[T1]{fontenc}

\usepackage{bm}

\newcommand{\be}{\begin{equation}}
\newcommand{\ee}{\end{equation}}

\begin{document}

\begin{center}
\vspace{24pt}
{ \large \bf 2d CDT with gauge fields}

\vspace{30pt}

{\sl J. Ambj\o rn}$\,^{a,b}$
and 
{\sl A. Ipsen}$\,^{a}$

\vspace{48pt}
{\footnotesize

$^a$~The Niels Bohr Institute, Copenhagen University\\
Blegdamsvej 17, DK-2100 Copenhagen \O , Denmark.\\
email: ambjorn@nbi.dk, asgercro@fys.ku.dk\\

\vspace{10pt}

$^b$~Institute for Mathematics, Astrophysics and Particle Physics (IMAPP)\\ 
Radbaud University Nijmegen, Heyendaalseweg 135,
6525 AJ, Nijmegen, The Netherlands

}
\vspace{96pt}
\end{center}


\begin{center}
{\bf Abstract}
\end{center}

\noindent 
Two-dimensional Causal Dynamical Triangulations  provides
a definition of the path integral for
projectable two-dimensional Horava-Lifshitz
quantum gravity. We solve the theory coupled to gauge fields.

\vspace{12pt}
\noindent

\vspace{24pt}

\newpage

\section{Introduction}

Two-dimensional causal dynamical triangulations (CDT)  \cite{al}
provides a well defined path integral representation
of two-dimensional projectable Ho\v{r}ava-Lifshitz quantum gravity 
(HL \cite{hl}),
as was recently shown in \cite{agsw}. 2d CDT coupled to conformal 
field theories with central charges $c=1/2$ and $c=4/5$ as well as
$c\geq 1$  have been investigate numerically \cite{aal,aal1,gauss}.
However, it has not yet  been possible to provide exact solutions
of the gravity theory coupled to a well defined continuum matter theory
despite the existence of a matrix formulation \cite{cdtmatrix}\footnote{To 
be precise, CDT has been solved when coupled to some ``non-standard''
hard dimer models \cite{aggs,az}, but it is unknown if these dimer models have 
an interesting continuum limit. Also, `` generalized CDT'' models coupled to 
ordinary hard dimer models have been solved \cite{aggs,az1}, 
using matrix models.}. Here we will provide
a first such step and solve CDT coupled to gauge theories. 

Gauge theories are simple in two dimensions since there 
are no propagating field degrees of freedom. However, if the 
geometry is non-trivial there can still be non-trivial 
dynamics, involving a finite number of degrees of freedom.
In the CDT case we consider space-time with the topology
of a cylinder, space being compactified to $S^1$, and we 
thus have non-trivial dynamics associated with the holonomies
of $S^1$. This has been studied in great detail in flat space-time
(see \cite{2dgauge} and references therein). 
We will use the results from these studies
to solve CDT coupled to gauge theory. The rest of this 
article is organized in the following way. In Sec.\ \ref{review}
we review the part of \cite{2dgauge} that we need for the
construction the CDT quantum Hamiltonian. In Sec.\ \ref{H} we 
find the lattice transfer matrix and the corresponding continuum
Hamiltonian and finally in  Sec.\ \ref{cosmo} we discuss 
``cosmological'' applications.

\section{2d gauge theories on  a cylinder}\label{review}

Let us first heuristically understand the Hamiltonian for 
gauge theory on the cylinder, the gauge group $G$ being a 
simple Lie group (we can think of $G=SU(N)$ if needed). The action is
\beq\label{j1}
S_{YM}= \frac{1}{4} \int dtdx \;(F^a_{\m\n})^2,~~~~ \m,\n=0,1,
\eeq
where $F^a_{01}=E_1^a$ is the chromo-electric field. Quantizing in
the temporal gauge, $A_0^a=0$, say, one obtains the Hamiltonian
\beq\label{j2}
\hH = \oh \int dx \;  (\hE_1^a)^2, 
~~~\hE_1^a\equiv -i \frac{\del}{\del A^a_1(x)},
\eeq
and this Hamiltonian acts on physical states, i.e.\ wavefunctions 
which satisfy Gauss law
\beq\label{j3}
(D_1 \hE^1)^a\Psi(A) =0,
\eeq
where $D_1$ denotes the covariant derivative. 
Since $D_1E^1$ are the generators of gauge transformations \rf{j3}
just tells us that $\Psi(A)$ is gauge invariant. But on $S^1$ the  
only gauge invariant functions are class functions of 
the holonomies and any class function can be expanded in 
characters of irreducible unitary representations of the group. 
Let $T_R^a$ denote the Lie algebra generators of the representation $R$,
where $\tr_R T_R^aT_R^a= C_2(R) $, the value of quadratic Casimir 
for the representation $R$. For a holonomy
\beq\label{j4}
U_R(A) = P \e^{i g \oint dx A^a_1(x)T_R^a}, ~~~\chi_R (U) \equiv \tr_R U_R,
\eeq
where $g$ is the gauge coupling, one easily finds that the action 
of $\hH$  on the wavefunction $\chi_R(U(A))$ is 
\beq\label{j5}
\hH \chi_R(U(A)) =\oh g^2 L C_2(R) \chi_R(U(A)).
\eeq 
From this we take along that on 
the gauge invariant wave functions we can  write\footnote{\label{foot1} We are 
clearly not very precise here when discussing the quantization
(that is why we started this section with the word ``heuristic''). 
We have still available a time independent gauge transformation
which we can use to gauge the holonomy $U(A)$ to a Cartan 
subalgebra of $G$, i.e.\ to diagonalize $U(A)$, and further 
to permute the diagonal elements. Strictly speaking $\hH$ 
should be defined on this subspace which is the orbifold
$T^{N-1}/S_N$ for $G=SU(N)$. We refer the reader to 
\cite{2dgauge} for details.} 
\beq\label{j6}
\hH = \oh g^2 L \Del_G
\eeq
where $\Del_G$ is the Laplace-Beltrami operator on the group $G$
(here $SU(N)$), and further that the gauge invariant eigenfunctions
are the irreducible characters of $G$.

\begin{figure}[t]
\begin{center}
\includegraphics[width=14cm]{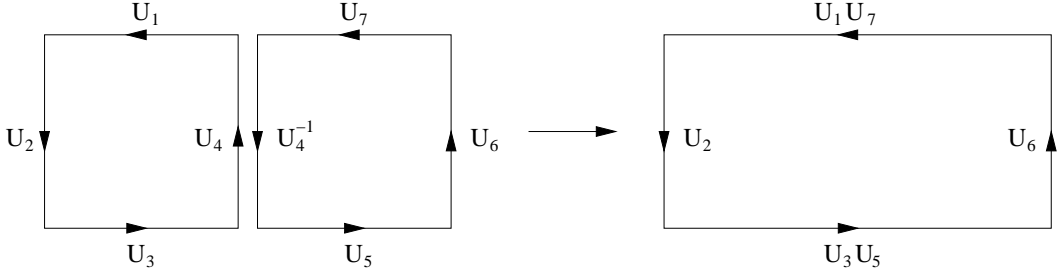}
\end{center}
\vspace{-5mm}
\caption{Integrating out the link $U_4$ using the heat kernel
action. The graphic notation is such one has cyclic matrix multiplication
on loops and if an arrow is reversed (oriented link $\ell \to -\ell$)
then $U_{-\ell} = U^{-1}_\ell$.}
\label{fig1}
\end{figure}

Let us now quantize the theory using a lattice, i.e.\
using a (regularized) path integral. The lattice partition function
is defined as 
\beq\label{j7}
Z(g) = \int \prod_\ell dU_\ell \prod_{{\rm plaquettes}} Z_P[U_P],
\eeq
where we to each link $\ell$ associate a $U_\ell \in G$, and $U_P$ is 
the product of the $U_\ell$'s around the plaquette (since we always take 
the trace of $U_P$ it does not matter which link is first in the product
as long as we keep the orientation). One writes $U_\ell = e^{a g iA_\ell^bt^b}$
where $\ell$ signifies a link in direction $\m=0$ or $\m=1$, $a$ is
the length of a lattice link and we choose $t^a$ to be generators of 
the Lie algebra of $G$ in the fundamental representation, normalized 
to $\tr t^b t^c = 1/2$. This establishes a formal relation
between the gauge fields $A_\ell$ and the group variables $U_\ell$.
One has a large choice for $Z_P[U_p]$, the only requirement being 
that $Z(g)$ in \rf{j7} should formally become the continuum 
path integral when the lattice spacing is taken to zero. Often 
the so-called Wilson action is used where 
\beq\label{j8}
Z_P[U_P] = \e^{\bt \tr (U_p+U_P^{-1})},~~~~\bt = \frac{1}{4g^2 a^2}.
\eeq
In the limit where $a\to 0$ one has 
$ \tr (U_p+U_P^{-1}) \to  1-a^4 g^2(F_{\m\n}^b)^2 +0(a^6)$, thus leading
to the correct naive continuum limit in \rf{j7} if $\bt$ scales as 
in \rf{j8}. For the purpose of extracting the Hamiltonian 
it is convenient for us 
to use a different $Z_P[U_p]$, the so-called heat kernel action
\beq\label{j9}
Z_P[U_P] = \la U_P| e^{-\oh g^2 A_P \Del_G}|I\ra =  \sum_R d_R \chi_R(U_P) \, 
\e^{-\oh g^2 A_P C_2(R)},
\eeq
where $A_P = a_t a_s$ denotes the area of the plaquette with spatial 
lattice link length $a_s$ and time-like link length $a_t$ (we will 
usually think of $a_s=a_t$), $I$ denotes 
the identity element in $G$ and, as above $\Del_G$ the Laplace-Beltrami 
operator on $G$. Using $U_\ell = (I + a g A_\ell^at^a \cdots)$ 
in the limit $a\to 0$,
and $\sum_R d_R \chi_R(U_P) =\del(U_P-I)$, one can show
that the continuum Yang-Mills action is formally reproduced. 
The convenient property of the heat kernel action in 2d is that 
it is additive, i.e.\ if we integrate over a link in \rf{j7} the 
action is unchanged: write $U_{P_1}= U_4U_3U_2U_1$ and 
$U_{P_2} = U^{-1}_4U_7U_6U_5$, then 
\beq\label{j10}
\int dU_4 Z_{P_1} [U_{P_1}] Z_{P_2}[U_{P_2}] = Z_{P_1+P_2} [U_{P_1+P_2}],
\eeq
where $U_{P_1+P_2} = U_7U_6U_5U_3U_3U_1$, see Fig.\ \ref{fig1}.

The relation follows from the orthogonality of the group characters:
\beq\label{j11}
\int dU \; \chi_R (XU) \chi_{R'} (U^{-1} Y) = 
\frac{1}{d_R} \del_{RR'} \chi_R(XY).
\eeq

\begin{figure}[t]
\begin{center}
\includegraphics[width=12cm]{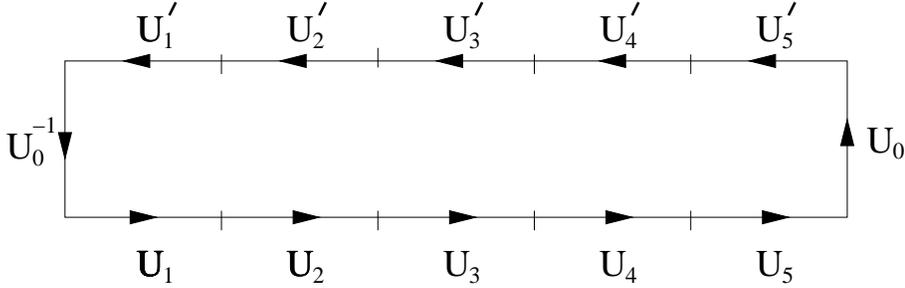}
\end{center}
\vspace{-5mm}
\caption{Integrating out the temporal links in a time-slab,
except a last link $U_0$. The temporal links $U_0^{-1}$ and $U_0$ 
are identified on the cylinder, 
and the result is $Z_P[U_P]$, 
$U_P=U_0(U_5U_4U_3U_2U_1)U_0^{-1}(U_1'U_2'U_3'U_4'U_5')$ 
using the heat kernel action. }
\label{fig2}
\end{figure}

Let us now consider a lattice with $t$ links in the time direction 
and $l$ links in the spatial direction. We have two boundaries,
with gauge field configurations $\{U_\ell\}$ and $\{U'_\ell\}$, which
we can choose to keep fixed (Dirichlet-like boundary conditions (BC)), 
integrate over (free BC),
or identify and integrate over (periodic BC). We write, using 
Dirichlet BCs
\beq\label{j12}
Z(g,\{U'_\ell\},\{U_\ell\}) = \la \{U'_\ell\}|\, \hT^t\,
|\{U_\ell\}\ra,~~~~\hT = \e^{-a_t \hH}
\eeq
where $\hT$ is the transfer matrix, giving us the transition 
amplitude between link configurations at neighboring
time slices. However in 2d we can restrict $\hT$ to be 
an operator only acting on the holonomies since  we can 
use \rf{j10} to integrate out the temporal   
links $U_\ell$ which connect two time slices. We obtain 
\beq\label{j13}
\la U'| \hT | U\ra = \int dU_0\; \la U'U_0 U^{-1} U_0^{-1}| \,
\e^{-l a_s a_t \oh g^2\Del_G} | e\ra,
\eeq  
where we have not integrated over the last temporal link $U_0$ and 
$U$ is the holonomy for the links at the time slice $t'$, say, and 
$U'$ the holonomy for the links at the neighbor  time slice $t'+1$
(see Fig. \ref{fig2}).

Using $\la U' U^{-1} | \e^{-\Del_G}|e\ra = \la U' | \e^{-\Del_G}|U\ra$ we 
can write the transfer matrix elements as 
\beq\label{j14}
\la U'|\, \e^{- a_t (l a_s \oh g^2\Del_G)} \hP |U\ra = 
\la U'|\hP\; \e^{- a_t (l a_s \oh g^2\Del_G)} \hP |U\ra,
\eeq
where the projection operator $\hP$ is defined by 
\beq\label{j15}
\hP |U\ra = \int dG\, |G U G^{-1}\ra.
\eeq
$\hP$  commutes with $\Del_G$, a fact  which allows us 
to write the right hand side 
of eq.\ \rf{j14} and thus  ensures that we can restrict the 
action of the transfer matrix even further, namely to the class functions
on $G$. To make this explicit consider an arbitrary state
\beq\label{j15a}
|\Psi \ra = \int dU \; |U\ra  \Psi(U),~~~~\Psi(U) = \la U|\Psi\ra,
\eeq
i.e.\ $(\hP \Psi)(U) = \int dG \Psi(G^{-1} U G) $ which is clearly a 
class function.

Denote the length of the lattice $L =a_s l$. From \rf{j12} 
and \rf{j14} it follows that
\beq\label{j16}
\hH = \oh g^2 L \Del_G.
\eeq
The expression agrees with the continuum expression.
We have reviewed  how the lattice theory, even 
if no gauge fixing is imposed, nevertheless makes it 
possible and natural to restrict the transfer matrix and 
the corresponding Hamiltonian to class functions of the holonomies.
Finally, it is of course only for the heat kernel action that
one derives an $\hH$ formally identical to the continuum 
Hamiltonian even before the lattice spacings are taken to zero.
The above arguments could be repeated for any reasonable action,
e.g.\ the Wilson action mentioned above, and in the limit where 
$a_s,a_t\to 0$ one would obtain \rf{j16}. Finally, the derivation 
can be repeated also for Abelian groups or discrete groups  like 
$Z_N$ groups,  resulting in an expression like \rf{j16}
with an appropriate group Laplacian $\Del_G$, in the Abelian case
without the issue of reduction of domain of $\Del_G$.

\section{Coupling to geometry}\label{H}
  
The covariant version of the Yang-Mills theory \rf{j1} is 
\beq\label{j17}
S_{YM} = \oq \int d^2 x \sqrt{g(x)}\; F_{\m\n}^a (F^{\m\n})^a.
\eeq
We want a path integral formulation which includes also 
the integration over geometries. Here the CDT formulation is 
natural: one is summing over geometries which have cylindrical
geometry and a time foliation, each geometry being defined by a 
triangulation and the sum over geometries in the path 
integral being performed by summing over all triangulations
with topology of the cylinder and a time foliation. The coupling 
of  gauge fields to a geometry via dynamical triangulations 
(where the length of a link  is $a$) is   
well known \cite{DTgauge}: One uses as plaquettes the triangles.
Thus the 2d partition function becomes
\beq\label{j18}
Z(\Lam,g,l',l,\{U'_\ell\},\{U_\ell\}) = 
\sum_{\cT} \e^{- \oh N_{\cT} \Lam \frac{\sqrt{3}}{4}a^2} Z^{G}_{\cT}(\bt),
\eeq
where the summation is over CDT triangulations $\cT$, 
with an ``entrance'' boundary consisting of $l$ links
and an ``exit'' boundary consisting of $l'$ links,  
$\Lam$ is the lattice cosmological constant, $N_{\cT}$ the 
number of triangles in $\cT$,  
and the gauge partition function for a given triangulation $\cT$
is defined as 
\beq\label{j19}
Z_{\cT}^{G}(g, \{U'_\ell\},\{U_\ell\}) = 
\int \prod_\ell dU_\ell \; \prod_P Z_P[U_P].
\eeq
The integration is over links and $\prod_P$ is the product 
over plaquettes (here triangles) in $\cT$.   For the plaquette 
action defining $Z_P[U_P]$ we have again many choices, and for 
convenience we will use the heat kernel action \rf{j9}.

We can introduce a transfer matrix $\hT$, which connects
geometry and fields at time label $t'$ to geometry and fields 
at time label $t'+1$, and if the (discretized) universe
has $t+1$ time labels we can write 
\beq\label{j20}
 Z(\Lam,g,l',l,\{U'_\ell\},\{U_\ell\}) = 
 \la \{U'_\ell\}, l' | T^t | \{U_\ell\},l\ra,~~~~T= \e^{-a \hH}.
\eeq
The one-dimensional geometry at $t'$ is characterized by 
the number $l$ of links (each of length $a$), and on these 
links we have field configurations $\{U_\ell\}$. Similarly
the geometry at $t'+1$ has $l'$ links and  field configurations
$\{U'_\ell\}$. For fixed $l$ and $l'$ the number of 
plaquettes (triangles) in the spacetime cylinder ``slab'' between $t'$ and 
$t'+1$ is $l+l'$ and the number of temporal links $l+l'$.
There is a number of possible triangulations of the slab for fixed 
$l$ and $l'$, namely
\beq\label{j21} 
N(l',l) =  \frac{1}{l+l'}\binom{l+l'}{l}.
\eeq
For each of these triangulations we can integrate over the 
$l+l'$ temporal link variables $U_0$, as we did for a fixed lattice
and we obtain as in that case
\beq\label{j22a}
\la U' |\hP\; \e^{-a (a (l+l') \frac{\sqrt{3}}{8} g^2\Del_G)} \hP |U\ra,
\eeq
where $U'$ and $U$ are the holonomies corresponding
to $\{U'_\ell\}$ and $\{U_\ell\}$, respectively, and $\hP$  
is the projection operator \rf{j15} to class functions coming from  
the last integration over a temporal link $U_0$. The factor $\sqrt{3}/8$
rather than the factor 1/2 appears because we are using equilateral 
triangles rather than squares as in Sec.\ \ref{review}. In order 
to have unified formulas we make a redefinition $g^2 \sqrt{3}/4 \to g^2$
and thus we have the matrix element
\beq\label{j22}
\la U' |\hP\; \e^{-a (a (l+l') \oh g^2\Del_G} \hP |U\ra,
\eeq 

If we did not have the matter fields the transfer matrix would be 
\beq\label{j23}
\la l'| \hT_{{\rm geometry}} |l\ra = N(l',l) \e^{-a((l+l')a \oh \Lam)},
\eeq 
where we have made a redefinition $\Lam \sqrt{3}/{4} \to \Lam$,
similar to the one made for $g^2$, in order to be in accordance 
with notations in other articles. 
The limit where $a \to 0$ and $L' =a \,l'$ and $L= a\,l$ 
are kept fixed has been studied \cite{2dcdt} and one finds
\beq\label{j23a}
\hT_{{\rm geometry}} = \e^{-a (\hH_{{\rm cdt}}+ O(a))},~~~~
\hH_{{\rm cdt}} = -\frac{d^2}{dL^2} L + \Lam L.
\eeq
From the definition \rf{j20} of $\hH$ and \rf{j22} it follows that
\beq\label{j24}
\hH =  \hH_{{\rm cdt}}  + \oh g^2 L \Del_G,
\eeq
acting on the Hilbert space which is the tensor product 
of the Hilbert space of square integrable class functions on $G$ and
the Hilbert space of the square integrable functions on $R_+$ with measure   
$d\m(L) = L dL$. 

We have obtained the Hamiltonian \rf{j24} using the path integral,
starting out with a lattice regularization. Alternatively one 
can use that the classical 2d YM action \rf{j1} on the (flat) cylinder 
can be formulated in terms of the holonomies  $U(t)$ defined in
eq.\  \rf{j4} (see \cite{rajeev} for details):
\beq\label{jx1}
 S_{YM} = \frac{1}{2} \int dx dt\; \tr E_1^2 = 
\frac{1}{2g^2 L} \int dt \;\tr (i U^{-1} \prt_0 U)^2.
\eeq
Let us now couple the YM theory to geometry as in \rf{j17}.
One observes that $\tilde{E}= E^1/\sqrt{g}=E_1 \sqrt{g}$ behaves as a scalar
under diffeomorphisms. Thus $D_1 \tilde{E} =0$, where $D_1$ is 
the ordinary gauge covariant derivative as in \rf{j3}. 
This implies that the derivation 
in \cite{rajeev} which led to \rf{jx1} for flat spacetime
is essentially unchanged. As we are interested in HL projectable
2d quantum geometries we assume the geometry is 
defined by a laps function $N(t)$, a shift function $N_1(x,t)$ 
and a spatial metric $\g(x,t)$. $\sqrt{g(x,t)} = N(t) \sqrt{\g(x,t)}$,
and introducing the spatial length $L(t) = \int dx  \sqrt{\g(x,t)}$
one obtains instead of \rf{jx1}
\beq\label{jx2} 
S_{YM} = \oh \int dt dx \; \sqrt{g(x,t)} \, \tr \tilde{E}^2 = \frac{1}{2g^2} 
\int dt \;\frac{\tr (i U^{-1} \prt_0 U)^2}{N(t) L(t)}.
\eeq  
Combined with the results from \cite{agsw} for the HL-action 
one can write the total action as 
\beq\label{jx3}
S_{TOT} =  \int dt \;\left[\frac{1}{2N(t) L(t)} 
\left(\oh (\prt_0 L)^2 +  
\frac{1}{g^2} \tr (i U^{-1} \prt_0 U)^2\right)+
{\Lambda} N(t) L(t)\right].
\eeq
This classical action leads to the quantum Hamiltonian \rf{j24}. 

Let us return to the quantum Hamiltonian \rf{j24}.
Since the eigenfunctions of $\Del_G$ after projection with $\hP$
are just the characters $\chi_R(U)$ on  $G$ 
and they have eigenvalues $C_2(R)$, we can solve the eigenvalue equation
for $\hH$ by writing $\Psi(L,U) = \psi_R(L) \chi_R(U)$.  
For $\hH_{{\rm cdt}}$ we have \cite{2dcdt,physrep} 
\beq\label{j25}
\hH_{{\rm cdt}} \psi_n (L,\Lam) = \ep_n \psi_n(L,\Lam), ~~~~~
\ep_n= 2n \sqrt{\Lam},~~n>0,
\eeq
where the eigenfunctions are of the form 
$\Lam \,p_n(L\sqrt{\Lam}) \e^{-\sqrt{\Lam} L}$, 
$p_n(x)$ being a polynomial of degree $n-1$.
The corresponding solution for $\psi_R(L)$ is obtained by the 
substitution 
\beq\label{j25a}
\Lam \to \Lam_R= \Lam + \oh g^2 C_2(R),
\eeq 
i.e.\
\beq\label{j26}
\hH \Psi_{n,R} = E(n,R) \Psi_{n,R},~~~E(n,R) = 2n  \sqrt{\Lam_R},~~n>0
\eeq
\beq\label{j27}
\Psi_{n,R}(L,U) = \Lam_R\, p_n(L \sqrt{\Lam_R})\, \e^{-L  \sqrt{\Lam_R}} \,
\chi_R(U),
\eeq
with the reservation that the correct variable is not really 
the group variable $U$ but rather the conjugacy class corresponding 
to $U$. In the simplest case of $SU(2)$ the group manifold can 
be identified with $S^3$ and $\Del_G$ is the Laplace-Beltrami
operator on $S^3$. The conjugacy classes are labeled by the geodesic 
distance $\tht$ to the north pole and  the representations are labeled
by $R=j$ and we have\footnote{\label{foot2}
Using the lattice we have effectively 
performed a quantization using the fact that $SU(2)$ is a compact group.
However, as already mentioned in footnote \ref{foot1}, there are 
subtleties associated with the quantization, more precisely 
whether one chooses first to project to the algebra and 
quantized there, or first to quantized using the group 
variables and then project to the holonomies. We refer to \cite{2dgauge}
for a detailed discussion.}  
\beq\label{j28}
C_j = j(j+1),~~~~\chi_j (\tht)  = 
\frac{\sin (j+\oh)\tht}{\sin \oh \tht},~~~j=0, \oh, 1, \ldots
\eeq

As already mentioned the above results are also valid in 
simpler cases. If $G=U(1)$ where one has 
\beq\label{j29}
U(\tht) = \e^{i \tht},~~~~\Del_G = -\frac{d^2}{d \tht^2},
\eeq
\beq\label{j30}
C_n = n^2,~~~~\chi_n( \tht) = \e^{i n  \tht},~~~~n=0,\pm 1,\pm2, \ldots.
\eeq
and  if $G=Z_N$, the discrete cyclic group of order $N$,
\beq\label{j31}
U(k) = \e^{\frac{2\pi}{N} \, k},~~~~~
(\Del_G)_{k,k'} = \del_{k,k'+1}+\del_{k,k'-1}-2\del_{k,k'}, ~~~k=0,\ldots,N-1,
\eeq 
\beq\label{j32}
C_n = 2\left(1-\cos\left(\frac{2\pi}{N} \, n\right)\right)~~~~
\chi_n(k) = \e^{i \frac{2\pi n}{N}\,k},~~~n=0,1,\ldots,N-1.
\eeq

\section{The ground state of the universe}\label{cosmo}

In CDT the disk amplitude is defined as 
\beq\label{j33}
W_\Lam (L) = \int_0^\infty dt\;\la L |\, \e^{-t \hH_{\rm cdt}}|L'\to 0\ra.
\eeq
It is a version of the Hartle-Hawking wave function. One 
can calculate $W_\Lam(L)$ \cite{al}:
\beq\label{j34}
W_\Lam(L) = \frac{\e^{-\sqrt{\Lam}L}}{L}.
\eeq
This function satisfy 
\beq\label{j35}
\hH_{{\rm cdt}} W_\Lam (L) = 0,
\eeq
and one can view \rf{j35} as the Wheeler-deWitt equation. Formally
$W_\Lam(L) \propto \psi_0(L)$ in the notation used in eq.\ \rf{j25}, but it was 
not included as an eigenfunction in the listing in \rf{j25}
since it does not belong to the 
Hilbert space $L^2(R_+)$ with measure $LdL$.

\begin{figure}[t]
\begin{center}
\includegraphics[width=10cm]{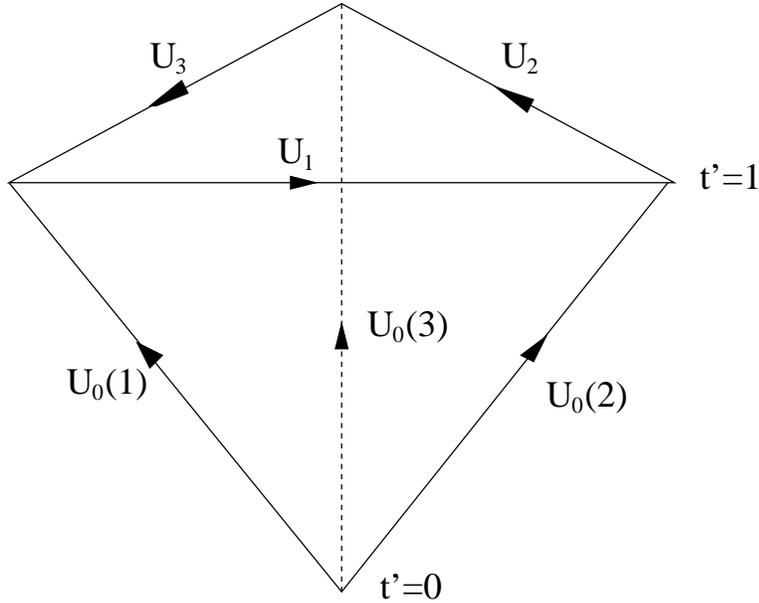}
\end{center}
\vspace{-5mm}
\caption{The ``beginning of the universe'' at $t'=0$ and 
the connection to the first loop at $t'=1$. }
\label{fig3}
\end{figure}

If we couple the theory of fluctuating geometries to gauge fields as above
we have to decide what kind of boundary condition to 
impose in the limit $L'\to 0$ in \rf{j33}.
A possible  interpretation of this  ``singularity'' 
in the discrete setting is that all the vertices  of
the first time slice at time $t'=1$ have additional temporal 
links joining  a single vertex  at time $t'=0$ (see Fig.\ \ref{fig3}).
We can view this as an explicit, discretized, realization of 
the matter part of the Hartle-Hawking boundary condition. 

Denote by  $\{U^{(0)}_\ell\}$, $\ell =1,\ldots,l$ the gauge fields on
these temporal links and  by $\{U_\ell\}$, $\ell =1,\ldots,l$ 
the gauge fields on the spatial links 
constituting the first loop at time $t'=1$ and denote by $U(1)$ 
the corresponding holonomy at time $t'=1$. 
The contribution to the matter partition function coming from 
this first ``big bang'' part of the universe is then 
\beq\label{j36}
\int \prod_{k=1}^{l} dU^{(0)}_k \; \prod_{k'=1}^l Z_{P_{k'}}[U_{P_{k'}}] =
Z_{{\rm disk}} [U(1)] = \la U(1)| \, \e^{-\oh g^2 l a^2 \Del_G} |I\ra ,
\eeq   
where we have integrated out the temporal links $\{U^{(0)}_\ell\}$.
The matter partition function can now be written (after integrating 
out the temporal links in the rest of the lattice too, as the integral 
over $t$ holonomies $U(1),U(2),\ldots,U(t)$
\beq\label{j37}
\int \prod_{i=1}^t \left( dU(i) \la U(i)| \, 
  \e^{-\oh g^2 (l_i+l_{i-1}) a^2 \Del_G}|U(i-1)\ra\right),
\eeq
where $U(0) \equiv I$ and $l_0=0$.
From this expression it is natural to say that the universe 
starts out in the matter state $|I\ra$, or expanded in charaters:
\beq\label{j38}
\la U| I\ra = \del(U-I) = \sum_R d_R \chi_R(U).
\eeq
This wave function is not normalizable if the group has 
infinitely many representations, but neither is $W_\Lam(L)$ as 
we just saw. Combining the two we might define the Hartle-Hawking wavefunction
for 2d CDT coupled to gauge fields as 
\beq\label{j39}
W(L,U) = \int_0^\infty dT \, \la L,U|\,\e^{-T \hH} |L=0,U=I\ra  
= \sum_R d_R \chi_R(U) \; W_{\Lam_R}(L),
\eeq
where $\Lam_R$ is defined in eq.\ \rf{j25a}.
We have explicitly: 
\beq\label{j40}
  W(L,k) = \sum_r \e^{\frac{i2\pi r k}{N}}  \;
\frac{\exp\left(-L (\sqrt{\Lambda + 
 g^2[ 1 -\cos(2\pi r/n)]})\right)}{L},
\eeq
for the $Z_N$ theory, 
\beq\label{j41}
  W(L,\theta) = 
\sum_{r=-\infty}^\infty \e^{ir\theta}\;
\frac{\exp \left(-L\sqrt{\Lambda+\oh r^2g^2 }\right)}{L}.
\eeq
for the $U(1)$ theory, and 
\beq\label{j42}
 W(L,\theta) = 
\sum_{k=0}^{\infty} 
\frac{\sin\left(\frac{(k+1)\tht}{2}\right)}{\sin \frac{\tht}{2}}\;
\frac{\exp \left(-L\sqrt{\Lambda+\frac{1}{8} g^2 k(k+2) }\right)}{L}.
\eeq
for the $SU(2)$ theory.

We have tried to define the initial matter state $|I \ra$ in the 
Hartle-Hawking spirit as coming from ``no boundary'' conditions
by closing the universe into a disk. Even if the ``initial'' 
(Big Bang) state is then a simple tensor product $|L=0\ra \otimes |I\ra$,
the corresponding Hartle-Hawking wave function is the result 
of a non-trivial interaction between matter and geometry. 
However, we cannot claim that the model 
points to such a ``no boundary'' condition in a really 
{\it  compelling} way. From a continuum point of view it should not 
make a difference if we, rather than implementing the continuum 
statement $L' \to 0$ by adding a little cap, had implemented it 
by insisting that the first time slice had $l=2$ or $l=3$, say.
The calculation of $W_\Lam(L)$ is insensitive to such details. However,
if our universe really started with such a microscopic loop, there 
is no reason that we should not choose the matter ground state,
i.e.\ the trivial, constant, character as the initial state. 
In this case absolutely nothing happens with matter during the time evolution 
of the universe. It just stays in this state and the state 
does not influence the geometry. Clearly the state $| I\ra$ is 
much more interesting and more in accordance with the picture 
we have of the Big Bang of the real 4d world where 
matter and geometry have interacted.  
Even if the argument for the state $|I\ra$ are not compelling, as 
just mentioned, it is nevertheless encouraging that the  
``natural'' Hartle-Hawking like boundary condition leads to a non-trivial 
interaction between geometry and matter.

\vspace{.5cm}      
\noindent {\bf Acknowledgments.} 
The authors acknowledge support from the ERC-Advance grant 291092,
``Exploring the Quantum Universe'' (EQU) as well as the  support 
from FNU, the Free Danish Research Council, from the grant 
``quantum gravity and the role of black holes''. Finally this research 
was supported in part by the Perimeter Institute of Theoretical Physics.
Research at Perimeter Institute is supported by the Government of Canada
through Industry Canada and by the Province of Ontario through the 
Ministry of Economic Development \& Innovation.

\end{document}